\documentclass{article}
\usepackage{amsmath,amssymb,amsfonts}
\usepackage{spconf,graphicx}
\usepackage{cite}
\usepackage{algorithmic}
\usepackage{textcomp}
\usepackage{xcolor}
\usepackage{bm}
\usepackage{pgfplots}
\usepackage{pgfplotstable}
\usepackage{subfig}
\usepackage{multirow}

\title{Compressing Flow Fields with \\
	Edge-aware Homogeneous Diffusion Inpainting}
\name{Ferdinand Jost \qquad Pascal Peter \qquad Joachim Weickert
	\thanks{This work has received funding from the European 
Research Council
		(ERC) under the European Union's Horizon 2020 
research and
		innovation programme (grant agreement no. 741215, ERC 
Advanced
		Grant INCOVID).}
	}

\address{Mathematical Image Analysis Group, Faculty of Mathematics 
and Computer Science\\
	Campus E1.7, Saarland University, 66041 Saarbr\"ucken, 
Germany\\
	\{jost, peter, weickert\}@mia.uni-saarland.de}

\begin{document}
%
\maketitle
\begin{abstract}
In spite of the fact that efficient compression methods for dense 
two-dimensional flow fields would be very useful for modern video 
codecs, hardly
any research has been performed in this area so far. Our paper
addresses this problem by proposing the first lossy diffusion-based
codec for this purpose. It keeps only a few flow vectors on a 
coarse grid. Additionally stored edge locations ensure the accurate 
representation of discontinuities. In the decoding step, the 
missing information is recovered by homogeneous diffusion 
inpainting that incorporates the stored edges as reflecting boundary
conditions. In spite of the simple nature of this codec, our 
experiments show that it achieves remarkable quality for compression 
ratios up to $800 : 1$.
\end{abstract}
%

\section{Introduction}\label{sec:introduction}

Flow fields are omnipresent in many applications ranging from 
scientific 
visualisation over object tracking to video compression.
Most modern video codecs such as HEVC~\cite{ISO:hevc} use block 
matching algorithms to compute motion vectors for coarse blocks of 
pixels. The resulting piecewise constant motion fields can be encoded 
very efficiently, but introduce block artefacts. 
Alternatively, optical flow methods have been used to compute dense 
flow fields~\cite{DBLP:conf/icip/KrishnamurthyMW95, 
DBLP:journals/tip/HanP01}
that give a better prediction, but are harder to encode efficiently.
Unfortunately, there are no established codecs that specialise on the 
compression of dense flow fields. Some notable attempts 
incorporate coding costs into optic flow estimation~\cite{DBLP:conf/cvpr/OttavianoK13} or
come up with an end-to-end framework for video coding with dense flow 
fields~\cite{lu2019dvc}. Another approach focuses on preserving the topology
of flow fields for scientific visualisation purposes~\cite{DBLP:journals/cgf/TheiselR03}. 

\subsection{Our Contribution}

The goal of our present paper is to close this gap by proposing a 
dedicated codec for flow fields. To this end we 
use ideas from diffusion-based inpainting of sparse data that respects 
discontinuities. These approaches have shown excellent performance 
for encoding cartoon-like still images~\cite{MBWF11} and piecewise smooth 
depth maps~\cite{HMWP13}. Although this strategy appears promising since 
flow fields are often piecewise smooth, it has never been tried before.
It requires a number of specific adapations along with a thorough
evaluation. We show that a carefully engineered combination of
simple ingredients such as Marr--Hildreth edge detection with 
hysteresis thresholding, homogeneous diffusion inpainting on a sparse 
regular grid, and uniform quantisation results in a codec that 
offers excellent performance for compression ratios up to $800 : 1$.



\subsection{Related Work}

Compression methods with diffusion-based inpainting were introduced 
by Gali{\'{c}} et al.~\cite{GWWB05} in 2005. Their method stores only 
a few selected pixels (so-called mask pixels) of an image and 
reconstructs the missing data with 
edge-enhancing anisotropic diffusion (EED)~\cite{We94e}. The \mbox{R-EED} 
algorithm of Schmaltz et al.~\cite{SPME14} improved this idea with  
a rectangular subdivision that adaptively inserts mask pixels. It
can beat the quality of JPEG2000 for some real-world test images.

The concept of diffusion-based inpainting was also extended to video 
compression. Peter et al.~\cite{PSMM15} developed a method based on 
R-EED that allows decoding in real time. However, this approach 
compresses each frame individually and does not exploit temporal 
redundancies. Andris et al.~\cite{APW16} proposed a proof-of-concept 
video codec that additionally uses optical flow methods for inter 
frame prediction. Their motion fields are compressed with a simple 
subsampling, again resulting in block artefacts.

The above mentioned techniques were tailored to natural data 
with a realistic amount of texture. This often requires fairly advanced
inpainting methods such as edge-enhancing anisotropic diffusion.
However, if one has piecewise smooth data that do not contain 
texture, simpler operators such as homogeneous diffusion inpainting
are sufficient, if they are supplemented with additional edge
information.

Mainberger et al.~\cite{MBWF11} developed such a codec for 
cartoon-like still images. It stores mask pixels on both sides 
of edges and uses homogeneous diffusion inpainting to reconstruct 
smooth regions in-between. Gautier et al.~\cite{GLG12} have extended
these ideas to depth maps. Hoffmann et al.~\cite{HMWP13} proposed
an alternative that explicitly stores segment boundaries, data
at a coarse hexagonal grid, and a few additional pixels at
locations where the reconstruction quality is poor. 

While this discussion shows that homogeneous diffusion inpainting 
which respects edges is adequate for piecewise smooth still images
and depth maps, its usefulness for compressing flow fields is an
open problem. This is the topic of our paper.
As we will see below, our approach also differs on a technical level
from the above mentioned papers: It does not require sophisticated
inpainting operators such as in \cite{GWWB05, We94e, SPME14, PSMM15}, 
does not use double-sided contour data as in \cite{MBWF11, GLG12}, 
and is substantially simpler than \cite{HMWP13}.

\subsection{Paper Structure}

In Section~\ref{sec:edgeinpainting} we design our edge-aware 
homogeneous diffusion inpainting operator for flow fields. With 
this concept we propose the encoding step of our method in 
Section~\ref{sec:encoding} and the corresponding decoding in 
Section~\ref{sec:decoding}. We evaluate our approach 
in Section~\ref{sec:experiments} and present our conclusions in 
Section~\ref{sec:conclusion}.

\section{Edge-aware Homogeneous Diffusion 
Inpainting}\label{sec:edgeinpainting}

The centrepiece of our codec is the \emph{edge-aware homogeneous 
diffusion} that reconstructs a flow field from a small amount of 
known data. This method relies on additional edge information to 
preserve discontinuities of the original flow field.
Consider a flow field $\bm{f}(\bm{x}): \Omega \rightarrow 
\mathbb{R}^2$ where $\bm{x} := (x,y)^\top$ denotes the position in a 
rectangular domain $\Omega \subset \mathbb{R}^2$. Flow vectors are 
only known at mask locations $K \subset \Omega$. We also assume 
that an edge set $E$ of $\bm{f}$ is given.
The reconstructed flow field $\bm{u}(\bm{x})$ can then be computed 
with homogeneous diffusion inpainting~\cite{Ii62} by solving the 
Laplace equation
\begin{equation}
\Delta \bm{u}(\bm{x}) = 0 \qquad\text{for}\quad \bm{x} \in 
\Omega \setminus (K \cap E)
\end{equation}
with fixed mask values and reflecting boundary conditions
\begin{align}
\bm{u}(\bm{x}) = \bm{f}(\bm{x}) &\qquad\text{for}\quad \bm{x} \in K \\
\partial_{\bm{n}} \bm{u}(\bm{x}) = 0 &\qquad\text{for}\quad \bm{x} 
\in \partial \Omega
\end{align}
where $\bm{n}$ denotes the normal vector to the boundary $\partial 
\Omega$.
As an extension to classical homogeneous diffusion we prevent any 
diffusion across discontinuities by introducing additional reflecting 
boundary conditions across the boundaries of $E$:
\begin{equation}
\partial_{\bm{n}} \bm{u}(\bm{x}) = 0 \qquad\text{for}\quad \bm{x} \in 
\partial E \, .
\end{equation}
This problem can be discretised with finite differences where edges 
are given at between-pixel locations. It yields a linear system of 
equations which we solve with a conjugate gradient method~\cite{Sa03}.
The result of this edge-aware homogeneous diffusion inpainting gives 
us a flow field with discontinuities along edges $E$ and smooth 
transitions in between.

\section{Encoding}\label{sec:encoding}

In this section we describe the encoder of our framework. First it 
collects edge information at between-pixel locations that 
act as boundaries for the homogeneous diffusion process. Then it 
selects the pixel mask needed for inpainting and quantises the 
selected pixel values. Finally, we use \emph{lpaq2}~\cite{Ma05} for 
entropy encoding.

\subsection{Edge Detection}

\begin{figure*}[ht]
	\centering
	\includegraphics[width=0.3\textwidth]{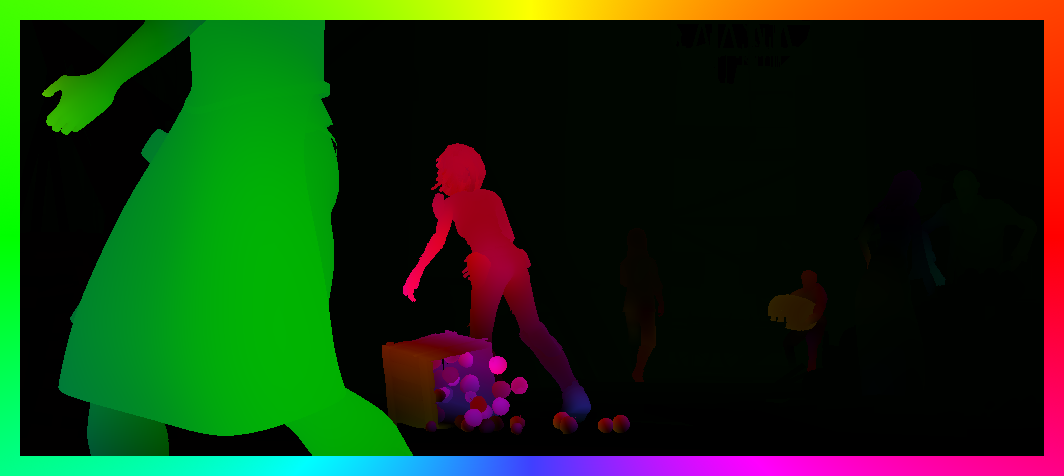}
	\includegraphics[width=0.3\textwidth]{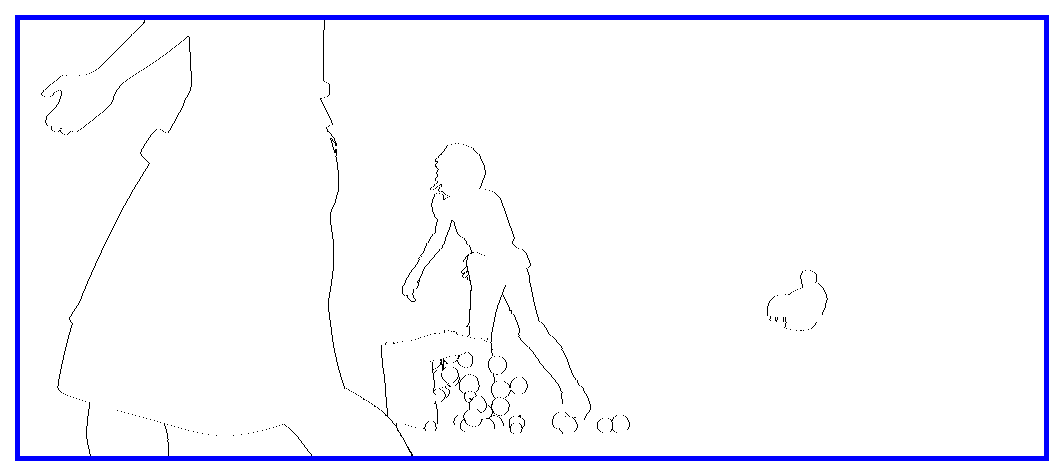}
	\includegraphics[width=0.3\textwidth]{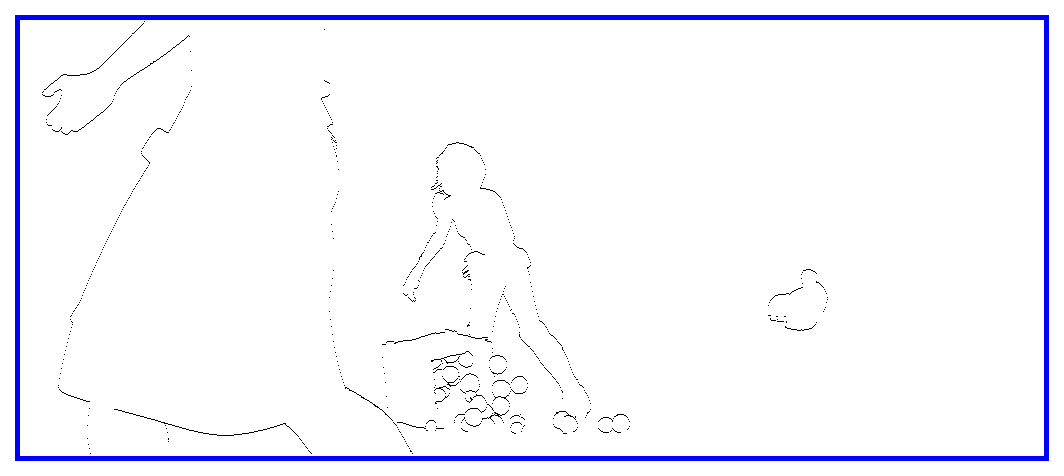}
	
	\caption{Edge detection with the Marr-Hildreth edge detector 
($\sigma = 0.5$, $T_1 = 4$, $T_2 = 2$). \textbf{Left:} Colour-coded 
original flow field. \textbf{Middle:} Edges in $x$-direction. 
\textbf{Right:} Edges in $y$-direction.}
	\label{fig:edges}
\end{figure*}

\begin{figure}[ht]
	\captionsetup[subfigure]{labelformat=empty}
	\centering
	\subfloat[type 1]{
		\begin{tikzpicture}[scale=0.4]
		\draw[step=1,gray,thin] (0.5,0.5) grid (3.5,3.5);
		\draw[line width=0.8mm, black] (2,1) -- (2,3);
		\draw[line width=0.8mm, black] (1,2) -- (2,2);
		\node at (1.5,2.5) [blue,circle,fill,inner 
		sep=1.5pt]{};
		\end{tikzpicture}
	}
	\subfloat[type 2]{
		\begin{tikzpicture}[scale=0.4]
		\draw[step=1,gray,thin] (0.5,0.5) grid (3.5,3.5);
		\draw[line width=0.8mm, black] (2,1) -- (2,3);
		\draw[line width=0.8mm, black] (2,2) -- (3,2);
		\node at (1.5,2.5) [blue,circle,fill,inner 
		sep=1.5pt]{};
		\end{tikzpicture}
	}
	\subfloat[type 3]{
		\begin{tikzpicture}[scale=0.4]
		\draw[step=1,gray,thin] (0.5,0.5) grid (3.5,3.5);
		\draw[line width=0.8mm, black] (1,2) -- (3,2);
		\draw[line width=0.8mm, black] (2,3) -- (2,2);
		\node at (1.5,2.5) [blue,circle,fill,inner 
		sep=1.5pt]{};
		\end{tikzpicture}
	}
	\subfloat[type 4]{
		\begin{tikzpicture}[scale=0.4]
		\draw[step=1,gray,thin] (0.5,0.5) grid (3.5,3.5);
		\draw[line width=0.8mm, black] (1,2) -- (3,2);
		\draw[line width=0.8mm, black] (2,1) -- (2,2);
		\node at (1.5,2.5) [blue,circle,fill,inner 
		sep=1.5pt]{};
		\end{tikzpicture}
	}
	\subfloat[type 5]{
		\begin{tikzpicture}[scale=0.4]
		\draw[step=1,gray,thin] (0.5,0.5) grid (3.5,3.5);
		\draw[line width=0.8mm, black] (2,3) -- (2,2);
		\node at (1.5,2.5) [blue,circle,fill,inner 
		sep=1.5pt]{};
		\end{tikzpicture}
	}
	\subfloat[type 6]{
		\begin{tikzpicture}[scale=0.4]
		\draw[step=1,gray,thin] (0.5,0.5) grid (3.5,3.5);
		\draw[line width=0.8mm, black] (1,2) -- (2,2);
		\node at (1.5,2.5) [blue,circle,fill,inner 
		sep=1.5pt]{};
		\end{tikzpicture}
	}
	
	\caption{Types of T-junctions with reference points in blue.}
	\label{fig:tjunctions}
\end{figure}
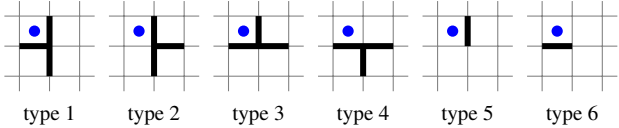

Our approach uses the Marr-Hildreth operator~\cite{MH80} combined 
with hysteresis thresholding similar to Canny \cite{Ca86} as an edge 
detector. It detects zero-crossings of the Laplacian of a 
Gaussian-smoothed image with a standard deviation of~$\sigma$. 
Additional hysteresis thresholding ensures that we keep only 
important structures: Edges where the gradient magnitude exceeds a 
threshold $T_1$ become seed points. Then we recursively add all 
candidates that are adjacent to seed pixels and exceed a threshold 
$T_2 < T_1$ to the set of edges.

The resulting relevant edges in $x$- and $y$-direction lie at 
between-pixel locations on two different grids. This gives us two 
binary images to encode (Fig.~\ref{fig:edges}). However,
following Hoffmann~et~al.~\cite{HMWP13}, we prefer an alternative 
that uses chain codes. This is very efficient in our setting as 
there are only three possible directions to follow.
We first extract different types of T-junctions as starting points 
for our chains (Fig.~\ref{fig:tjunctions}). Additional starting 
elements (types 5 and 6) are used to cover edges that do not contain 
any T-junctions. We then follow the contours until there is no more 
edge to encode. The chain code is a sequence of four different 
symbols: three symbols for each possible direction and a terminating 
symbol indicating the end of an edge.

\subsection{Inpainting Mask}

Our edge-aware homogeneous inpainting algorithm also requires a set 
of mask points to be able to reconstruct a flow field.
To reduce the coding cost of our method we choose mask pixels on a 
regular grid instead of arbitrary positions. Mask positions are 
uniquely determined by the image dimensions and a density parameter 
$d$.

Adaptive approaches like the rectangular subdivision of Schmaltz et 
al.~\cite{SPME14} usually give better results for diffusion processes. 
They accumulate mask pixels around edges to compensate the inability 
of homogeneous diffusion to preserve edges.  Our codec, however, 
already stores edge information explicitly. This way adaptive schemes 
do not provide a large benefit over a regular mask. In this case the 
small improvement in quality that these methods provide cannot 
compensate for the overhead of storing a more complex mask.

As edges can form isolated segments that do not contain a point of 
the regular grid, we additionally store the average flow value of 
such segments. For homogeneous diffusion,
this is equivalent to storing a single mask pixel at 
an arbitrary position inside this segment.

\subsection{Quantisation}

As flow vectors usually consist of two 32-bit floating-point numbers, 
we want to represent the range of actually occurring values as 
integers $q \in \{0,...,k-1\}$. One of the simplest methods to 
achieve this is \emph{uniform quantisation}.

Let $min$ be the minimal and $max$ the maximal value occurring in one 
channel of the flow field, and let $x \in [min, max]$ be a flow 
value. We also define the length of a quantisation interval as $a := 
\frac{max-min}{k-1}$.
The quantised value $q$ can then be computed as $q = \left\lfloor 
\frac{x-min}{a} + \frac{1}{2} \right\rfloor$.
In order to reconstruct a flow value $x_q$ we can simply compute $x_q 
= min + a \cdot q$.
The first and last interval only have a width of $\frac{a}{2}$. This 
preserves the minimal and maximal flow values.

\section{Decoding}\label{sec:decoding}

The decoding step of our codec is a straightforward process. We first 
reconstruct the regular inpainting mask using the stored density 
parameter $d$.
Next we follow the chain codes starting at T-junctions in each 
direction until we encounter a terminating symbol to reconstruct 
edges.
Finally, we restore flow vectors from the stored quantised values and 
place them on their corresponding mask positions.
The flow field is then reconstructed by solving the edge-aware 
homogeneous diffusion inpainting problem proposed in 
Section~\ref{sec:edgeinpainting}.

\section{Experiments}\label{sec:experiments}

\begin{figure*}[ht]
	\centering
	\setlength{\tabcolsep}{4pt}
	\begin{tabular}{c c c c c}

~ & \textbf{Original} & \textbf{Edge-aware} & \textbf{BPG} & \textbf{JPEG2000} \\
	
\multirow{1}{*}[6ex]{\rotatebox[origin=c]{90}{\textbf{image}}} &	
\includegraphics[width=0.22\textwidth]{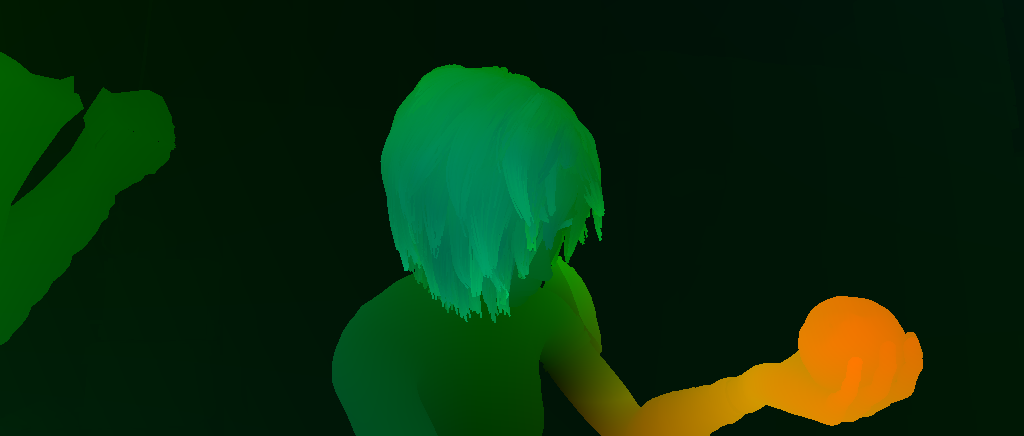} &
\includegraphics[width=0.22\textwidth]{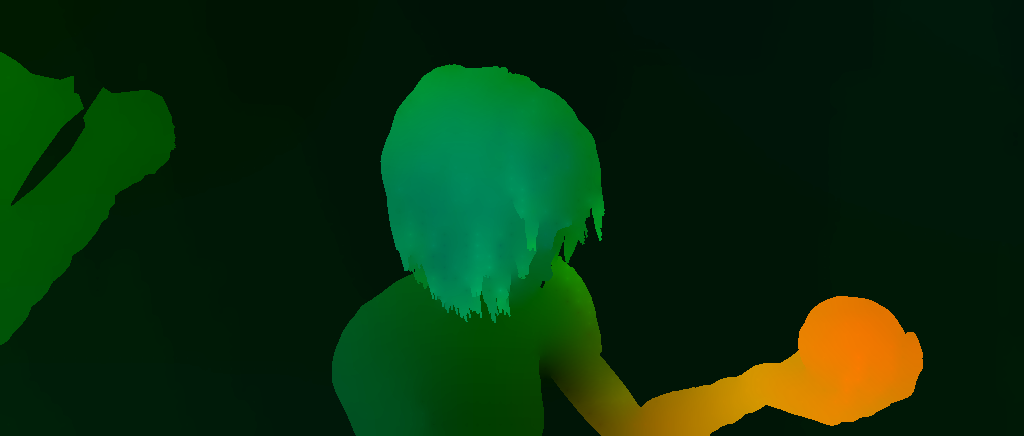} &
\includegraphics[width=0.22\textwidth]{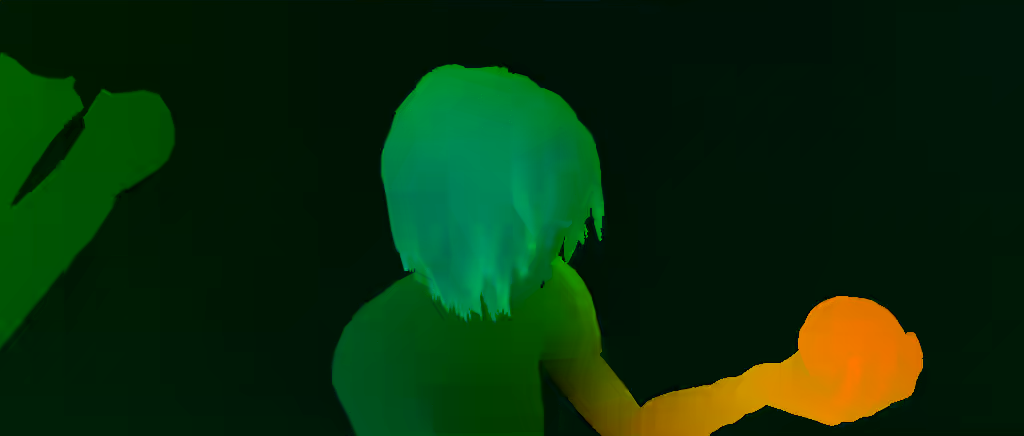} &
\includegraphics[width=0.22\textwidth]{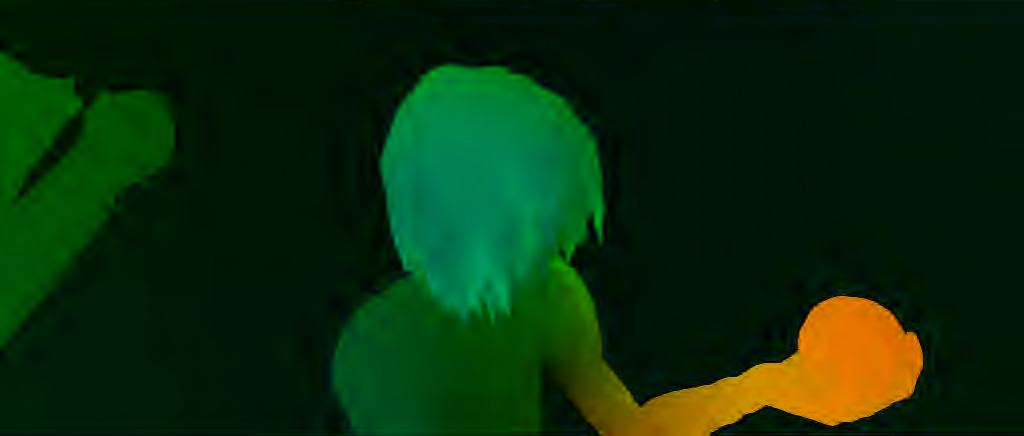} \\

\multirow{1}{*}[13ex]{\rotatebox[origin=c]{90}{\textbf{zoom}}} &
\includegraphics[width=0.22\textwidth]{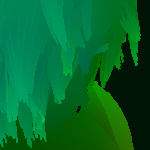} &
\includegraphics[width=0.22\textwidth]{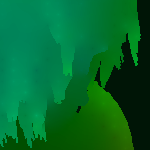} &
\includegraphics[width=0.22\textwidth]{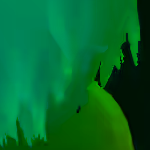} &
\includegraphics[width=0.22\textwidth]{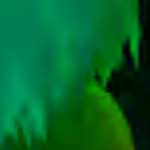} \\
		
 ~ & ~ & PSNR: $46.10$ & PSNR: $41.73$ & PSNR: $36.47$ \\

\\

\multirow{1}{*}[6ex]{\rotatebox[origin=c]{90}{\textbf{image}}} &
\includegraphics[width=0.22\textwidth]{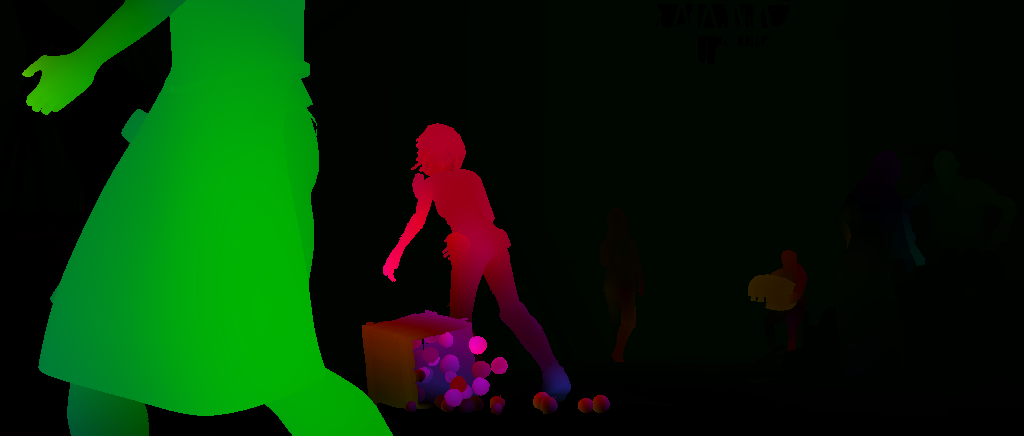} &
\includegraphics[width=0.22\textwidth]{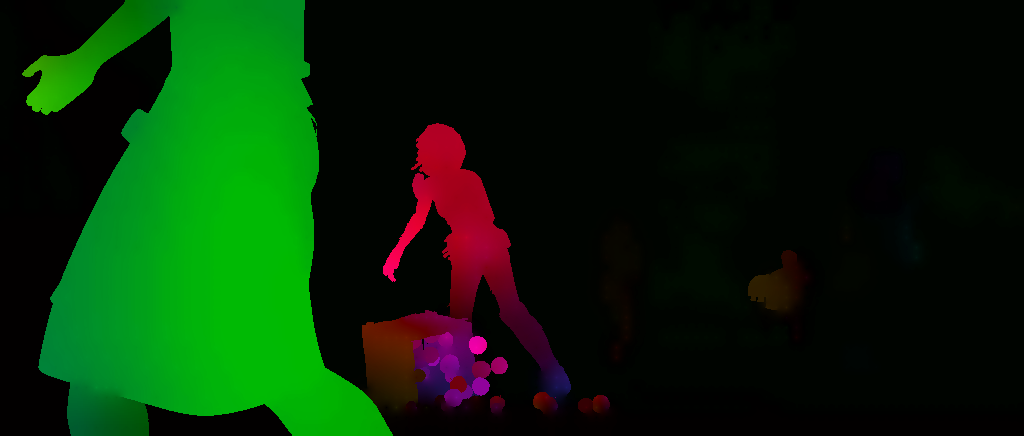} &
\includegraphics[width=0.22\textwidth]{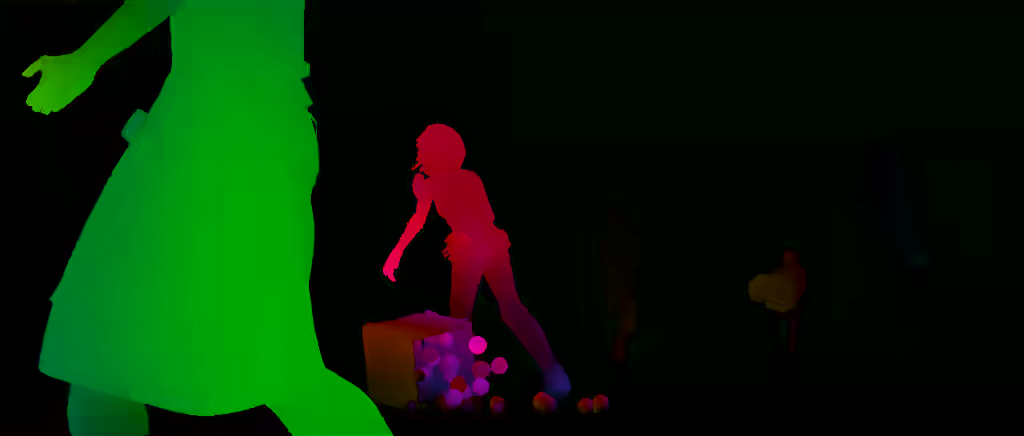} &
\includegraphics[width=0.22\textwidth]{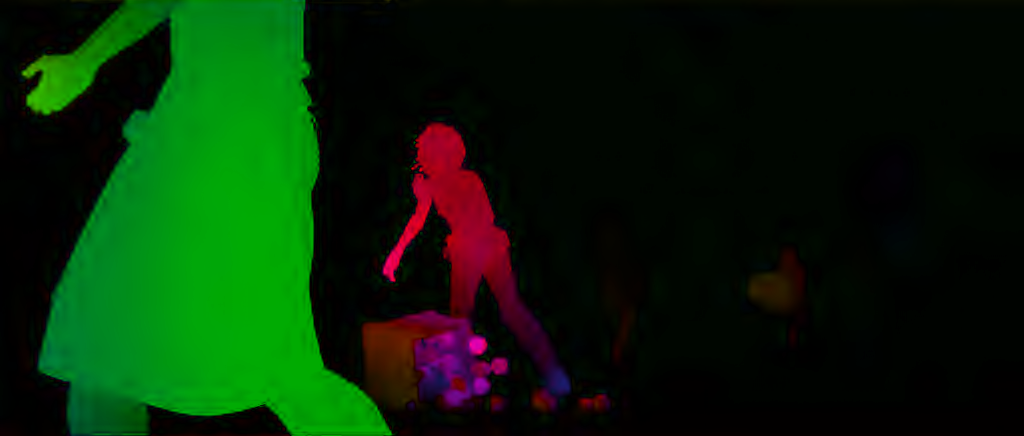} \\

\multirow{1}{*}[13ex]{\rotatebox[origin=c]{90}{\textbf{zoom}}} &
\includegraphics[width=0.22\textwidth]{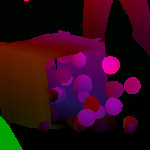} &
\includegraphics[width=0.22\textwidth]{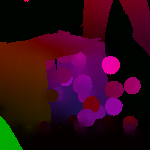} &
\includegraphics[width=0.22\textwidth]{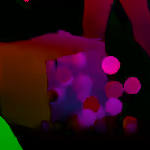} &
\includegraphics[width=0.22\textwidth]{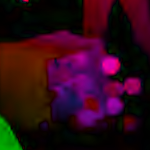} \\

~ & ~ & PSNR: $44.20$ & PSNR: $40.65$ & PSNR: $36.42$\\

	\end{tabular}
	
	\caption{Comparison of JPEG2000, BPG, and our edge-aware 
approach for two flow fields at a compression ratio of $400:1$. Both 
JPEG2000 and BPG show artefacts around edges, whereas our edge-aware 
approach preserves edges more accurately.}
	\label{fig:comparison}
\end{figure*}

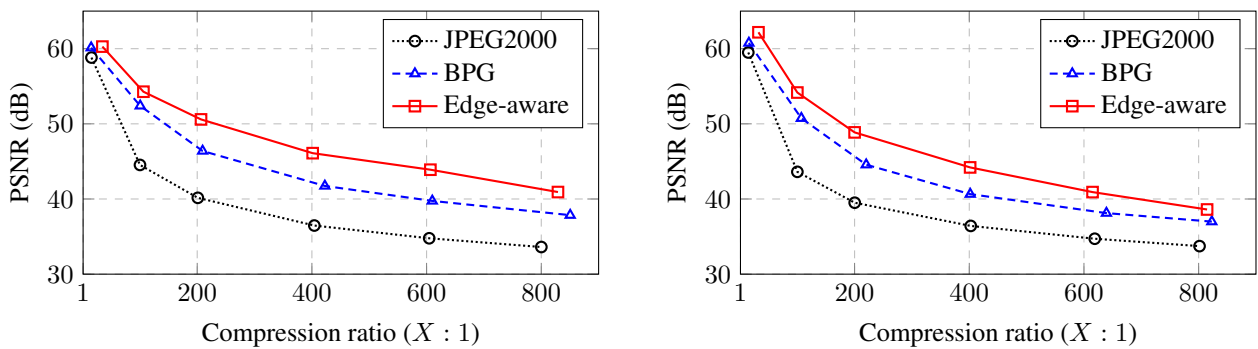
\begin{figure*}[!ht]
	\centering
	\begin{tikzpicture}[scale=1]
	\begin{axis}[
	/pgf/number format/.cd, 1000 sep={},
	name=plot1,
	y=0.60cm/6,
	xlabel={Compression ratio ($X:1$)},
	ylabel={PSNR (dB)},
	y label style={at={(axis description 
cs:0.1,.5)},anchor=south},
	xtick={1,200,400,600,800},
	ytick={30,40,50,60},
	xmin=1, xmax=900,
	ymin=30, ymax=65,
	legend pos=north east,
	ymajorgrids=true,
	xmajorgrids=true,
	grid style=dashed,
	legend cell align={left},
	scaled ticks=false, tick label style={/pgf/number 
format/fixed},
	]
	\addplot[color=black, mark=o, style=thick, densely dotted, 
mark options={solid}] table[x index=0,y index=1]{tab_alley_gt_grid};
	\addplot[color=blue, mark=triangle, style=thick, densely 
dashed, mark options={solid}] table[x index=2,y 
index=3]{tab_alley_gt_grid};
	\addplot[color=red, mark=square, style=thick] table[x 
index=6,y index=7]{tab_alley_gt_grid};
	\legend{JPEG2000\\BPG\\Edge-aware\\}
	\end{axis}
	
	\end{tikzpicture}
	\qquad
	\begin{tikzpicture}[scale=1]
	\begin{axis}[
	/pgf/number format/.cd, 1000 sep={},
	name=plot1,
	y=0.60cm/6,
	xlabel={Compression ratio ($X:1$)},
	ylabel={PSNR (dB)},
	y label style={at={(axis description 
cs:0.1,.5)},anchor=south},
	xtick={1,200,400,600,800},
	ytick={30,40,50,60},
	xmin=1, xmax=900,
	ymin=30, ymax=65,
	legend pos=north east,
	ymajorgrids=true,
	xmajorgrids=true,
	grid style=dashed,
	legend cell align={left},
	scaled ticks=false, tick label style={/pgf/number 
format/fixed},
	]
	\addplot[color=black, mark=o, style=thick, densely dotted, 
mark options={solid}] table[x index=0,y index=1]{tab_market_gt_grid};
	\addplot[color=blue, mark=triangle, style=thick, densely 
dashed, mark options={solid}] table[x index=2,y 
index=3]{tab_market_gt_grid};
	\addplot[color=red, mark=square, style=thick] table[x 
index=6,y index=7]{tab_market_gt_grid};
	\legend{JPEG2000\\BPG\\Edge-aware\\}
	\end{axis}
	
	\end{tikzpicture}
	
	\caption{Comparison of our method with JPEG2000 and BPG in 
terms of PSNR (higher is better). Our codec outperforms JPEG2000 and 
BPG consistently on both \emph{alley} (left) and \emph{market} 
(right).}
	\label{fig:graphs}
\end{figure*}

In this section we demonstrate the potential of our framework using two 
ground truth flow fields of the MPI Sintel Flow 
Dataset~\cite{DBLP:conf/eccv/ButlerWSB12} as test images. This provides 
us with high quality flow fields with realistic complexity. Since there 
are no standards for the compression of flow fields, we compare our method 
to JPEG2000~\cite{TM02} and BPG~\cite{website/bpg}, an adaption 
of HEVC's intra-coding mode for still images~\cite{DBLP:conf/pcs/NguyenM12}. 
Those are the best established alternatives for arbitrary images,
but are not designed to handle floating-point pixel data. 
Therefore, we quantise the input flow fields channelwise to integer values 
in the range $\{0,...,255\}$ with the uniform quantisation described in 
Section~\ref{sec:encoding}. We also use these quantised flow fields 
as inputs to our method.
To optimise the parameters of our approach we perform a simple grid 
search. We fix the standard deviation $\sigma$ of the edge detection 
to $0.5$, which performs well in our experiments.

Fig.~\ref{fig:comparison} shows the results of JPEG2000, BPG, and our 
edge-aware approach for a compression ratio of $400:1$. The 
compression ratio is computed relative to the quantised input flow 
fields. We observe that transform-based codecs like JPEG2000 have 
problems preserving 
sharp edges. This results in unpleasant ringing artefacts around 
object boundaries. BPG performs much better than JPEG2000, but still 
blurs edges. In contrast, our edge-aware homogeneous diffusion 
inpainting reconstructs all important edges very well.
In a quantitative comparison we measure the peak-signal-to-noise 
ratio (PSNR) of the different approaches. 
Fig.~\ref{fig:graphs} shows that our method outperforms both JPEG2000 
and BPG for all measured compression ratios, ranging from $10 : 1$ to $800 : 1$.

\section{Conclusions and Outlook}\label{sec:conclusion}

In our paper, we have presented the first diffusion-based compression
approach for dense flow fields. We have shown that a careful combination
of very simple concepts ranging from Marr-Hildreth edge detection
to uniform quantisation is already sufficent to achieve a remarkable
quality even at very high compression ratios. Our proof-of-concept 
framework is very flexible: Its modular nature allows to adapt and 
improve it by simply exchanging the different components. 
In our ongoing work, we are investigating this option further.
We will also embed our approach into existing video codecs.

\pagebreak

\interlinepenalty=10000
\bibliographystyle{IEEEbib}
\bibliography{strings,refs,myrefs}

\end{document}